\documentclass[twocolumn,showpacs,preprintnumbers,amsmath,amssymb]{revtex4}

\usepackage{graphicx}
\usepackage{dcolumn}
\usepackage{bm}

\begin{document}


\title{Low-Temperature Hall Effect in Substituted Sr$_{2}$RuO$_{4}$}

\author{Naoki Kikugawa, Andrew Peter Mackenzie}
\affiliation{
School of Physics and Astronomy, University of St.~Andrews, St.~Andrews, Fife KY16 9SS, United Kingdom
}
\author{Christoph Bergemann}
\affiliation{
Cavendish Laboratory, University of Cambridge, Madingley Road, 
Cambridge CB3 0HE, United Kingdom
}
\author{Yoshiteru Maeno}
\affiliation{
International Innovation Center, Kyoto University, Kyoto 606-8501, Japan\\
Department of Physics, Kyoto University, Kyoto 606-8502, Japan
}

\date{\today}

\begin{abstract}

We report the results of a study of the Hall effect and magnetoresistance 
in single crystals of Sr$_{2}$RuO$_{4}$ in which 
Sr$^{2+}$ has been substituted by La$^{3+}$ 
(Sr$_{2-y}$La$_{y}$RuO$_{4}$) or Ru$^{4+}$ by Ti$^{4+}$ 
(Sr$_{2}$Ru$_{1-x}$Ti$_{x}$O$_{4}$).  
For undoped Sr$_{2}$RuO$_{4}$, 
the purity is so high that 
the strong-field Hall coefficient can be measured for fields above 4\,T. 
The conventional weak-field Hall coefficient as a function of doping 
shows a sharp jump and sign change at $y$\,$\sim$\,0.01 that 
is unrelated to either a sharp change in Fermi-surface topography 
or a magnetic instability.
The implications of these results are discussed. 

\end{abstract}

\pacs{74.70.Pq, 74.62.Dh, 74.25.Fy}

\maketitle

\section{Introduction}

Transport properties under magnetic field such 
as Shubnikov-de Haas oscillations, magnetoresistance, 
and Hall resistivity reflect the Fermi-surface topology in metals \cite{Hurd}. 
In principle, 
the Hall effect provides information regarding 
both the orbital contribution from quasi-particles tracing the Fermi surface 
and the magnetic contribution (the ''anomalous'' Hall effect). 
It is therefore 
a widely-used probe in the investigation of highly correlated electron systems, 
where the ground states are dominated by interactions 
among $d$- and $f$-electrons \cite{Lee,Nakajima}.  
Recently, 
there has been specific interest in using it 
as a probe for the Fermi-volume changes that 
have been predicted to occur near some magnetic quantum critical points \cite{Coleman,Norman,Yeh}. 
Although it is commonly studied, 
the Hall effect is a difficult experiment to interpret. 
Separating the anomalous and orbital parts is not always easy, 
and even when this can be done, 
extracting information from the orbital part has 
a number of important complications. 
In the commonly-used weak field limit 
($\omega_{\rm c}\tau$\,$\ll$\,1, $\omega_{\rm c}$: band-averaged cyclotron frequency, 
$\tau$: band-averaged relaxation time), 
the Hall conductivity is a weighted Fermi-surface integral 
involving the square of the mean free path $\ell$, 
biased towards regions of high curvature. 
At high temperatures, 
where $\ell$ is expected to be $k$-dependent, 
it is hard to use the weak field Hall effect 
to obtain quantitative volume information on even simple Fermi surfaces. 
For multi-band materials, 
extracting quantitative information is next to impossible in this regime.
There is, however, 
a limit in which more quantitative analysis should in principle be possible. 
At very low temperatures, where elastic scattering dominates, 
$\ell$ is assumed to be $k$-independent, 
in the so-called ''isotropic-$\ell$'' limit \cite{Ong}. 
In this limit, 
the Hall conductivity is dominated by curved parts of the Fermi surface. 
In two dimensions, an analysis introduced by Ong \cite{Ong} yields 
a simple expression for Hall coefficient $R_{\rm H}$ 
for a multi-band material \cite{Andy_Hall}: 
\begin{eqnarray}
R_{\rm H} = \frac{2\pi d \sum_{i}(-1)^{n_{i}} \ell_{i}^{2}}{e(\sum_{i}k_{\rm F}^{i} \ell_{i})^{2}}
\label{eq1}. 
\end{eqnarray}
Here, 
$k_{\rm F}^{i}$ is the average Fermi wave-vector of the $i$-th Fermi-surface sheet, 
$\ell_{i}$ the mean free path of the quasi-particle on that sheet, 
$d$ the distance between adjacent RuO$_{2}$ layers, 
$n_{i}$\,=\,0 if $i$ is a hole pocket and 1 if it is an electron one. 
If $\ell$ is isotropic, 
it is independent of the sheet  
and disappears from Eq.~(\ref{eq1}), 
allowing an analysis of the Hall coefficient purely 
in terms of the Fermi-surface topography.  
Inspection of Eq.~(\ref{eq1}) shows that 
although it is possible to calculate $R_{\rm H}$ 
on the basis of known Fermi-surface topography for a material 
with more than one Fermi-surface sheet, 
the inverse is not in general true. 
However, 
that is not a very important constraint on the use of the Hall effect 
in studies of quantum criticality, 
because the key prediction is that of a sudden change in $R_{\rm H}$ 
at the critical points as a tuning parameter is varied \cite{Coleman,Norman}. 
One of the questions that we address with the experiments reported here is 
whether a sudden change in Fermi-surface topography is 
the only physics that can give a sudden change in $R_{\rm H}$ 
in a multi-band material as a function of doping in the low temperature limit. 
In the course of our work, 
we have also studied the Hall effect and 
magnetoresistance in the regime $\omega_{\rm c}\tau$\,$>$\,1, 
thus making one of the first observations, 
to our knowledge, 
of the strong-field Hall coefficient of a correlated electron material.
For our study, 
we concentrate on the two-dimensional metal and 
unconventional superconductor \cite{Maeno_Nature,Andy_Review}, Sr$_{2}$RuO$_{4}$, 
because the evolution of its Fermi surface has been traced by direct experimental measurement of the de Haas-van Alphen effect 
\cite {Andy_dHvA,Bergemann_Review}, 
even in the presence of chemical doping \cite{Kiku_Rigid}. 
Using the known Fermi-surface parameters, 
it was possible for ''pure'' Sr$_{2}$RuO$_{4}$ to obtain accurate agreement 
between the value of $R_{\rm H}$ calculated using Eq.~(\ref{eq1}) 
with the isotropic-$\ell$ approximation and direct measurement \cite{Andy_Hall}. 
A later Hall effect study of Ca-substituted Sr$_{2}$RuO$_{4}$ showed 
a strong dependence on the Ca concentration 
which was assumed to reflect a changing Fermi-surface topography 
coupled with the known changes in the crystal structure \cite{Galvin}. 
It was not, however, possible to prove that 
the Hall coefficient changes with doping were the result purely of 
a change in the Fermi-surface topography, 
because the high levels of disorder in the samples 
precluded independent measurement by quantum oscillations.
%

\begin{figure*}
\includegraphics[width=100mm]{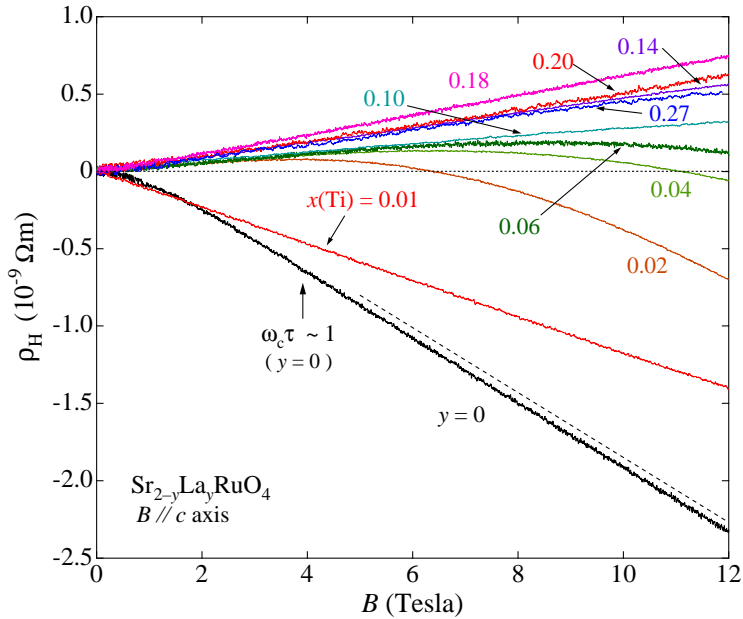}
\caption{\label{Fig1} 
The magnetic field dependence of the Hall resistivity $\rho_{\rm H}$ 
in Sr$_{2-y}$La$_{y}$RuO$_{4}$. 
The dashed line emphasizes the linear field dependence of $\rho_{\rm H}$ 
at high fields in pure Sr$_{2}$RuO$_{4}$ 
with $\rho_{ab0}$\,=\,90\,n$\Omega$cm. 
For comparison, 
the Hall resistivity in Sr$_{2}$Ru$_{1-x}$Ti$_{x}$O$_{4}$ 
with $x$\,=\,0.01 is also shown. 
It is linear over the whole field range, 
because $\rho_{ab0}$\,=\,5\,$\mu\Omega$cm and so  
only the weak-field regime can be accessed in fields up to 12\,T. 
}
\end{figure*}

%
Recently, 
the opportunity has arisen to check the behavior of 
the low temperature Hall effect in doped samples of Sr$_{2}$RuO$_{4}$ 
in a highly controlled situation. 
If La$^{3+}$ is substituted for Sr$^{2+}$ in Sr$_{2-y}$La$_{y}$RuO$_{4}$, 
the electrons that are doped onto the conduction band 
give a change in the Fermi-surface topography. 
It was possible to observe quantum oscillations over the range 
0\,$\leq$\,$y$\,$\leq$\,0.06, 
confirming that the doping resulted in a simple rigid-band shift of the Fermi level 
even in a strongly correlated, multi-band metal \cite{Kiku_Rigid}. 
Further analysis showed that 
the same rigid-band model gave a successful description 
of the measured specific heat to even higher values of $y$, 
motivating us to calculate the behavior of the Hall effect in the simplest model 
(rigid-band shift and isotropic-$\ell$) and then perform experiments 
to see the extent to which the measured Hall coefficient 
agreed with the prediction. 
Here we report the results of these measurements and a further data set 
acquired on single crystals of the Ti$^{4+}$ substituted system 
(Sr$_{2}$Ru$_{1-x}$Ti$_{x}$O$_{4}$). 
A further motivation for the work was studying the Hall effect in the presence of magnetic instabilities which previous work has shown 
to exist at $y$\,$\sim$\,0.20 \cite{Kiku_Band} and $x$\,$\sim$\,0.025 
\cite{Kiku_Ti}. 
%

\section{Experimental}

\begin{figure}
\includegraphics[width=65mm]{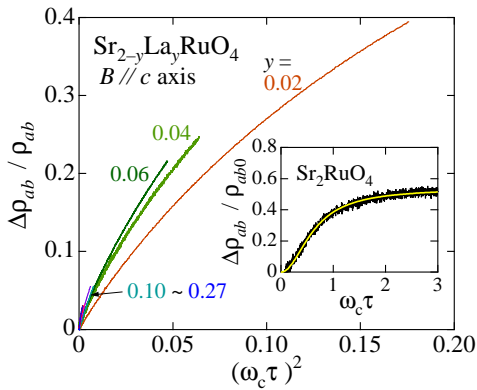}
\caption{\label{Fig2} 
Magnetoresistance of Sr$_{2-y}$La$_{y}$RuO$_{4}$ with $y$ up to 0.27 
as a function of $(\omega_{\rm c}\tau)^{2}$. 
The inset shows the magnetoresistance in pure Sr$_{2}$RuO$_{4}$ 
as a function of $\omega_{\rm c}\tau$. 
The curve running through the data is a fit using Eq.~(\ref{eq3}). 
}
\end{figure}

%
A series of single crystals of Sr$_{2-y}$La$_{y}$RuO$_{4}$ with $y$ up to 0.27 
and Sr$_{2}$Ru$_{1-x}$Ti$_{x}$O$_{4}$ with $x$ up to 0.09 were 
grown by a floating-zone method in an infrared image furnace 
(NEC Machinery, model SC-E15HD) in Kyoto, 
using procedures described in detail elsewhere 
\cite{Kiku_Band,Minakata}. 
The La and Ti concentrations in the crystals were determined by 
electron-probe microanalysis (EPMA). 
For the measurements described here, 
the crystals were cut into rectangles with a typical size of 2\,$\times$\,0.3\,$\times$\,0.03 mm$^{3}$, with the 
shortest dimension along the $c$ axis. 
Silver paste (Dupont, 6838) cured at 450\,$^{\circ}$C for 5\,minutes 
was used for attaching 8 electrodes onto each crystal in the Hall geometry; 
contact resistances were confirmed as below 0.3\,$\Omega$ 
at room temperature.   
Before measuring the Hall effect, 
we carefully checked the crystal homogeneity 
by multi-configuration resistivity measurements down to 4.2\,K 
in zero magnetic field. 
The temperature dependences of resistivity and residual resistivities 
$\rho_{ab0}$ reproduce well the results of previous studies \cite{Kiku_Rigid,Kiku_Ti,Kiku_JPSJ,Kiku_Band}. 
Here, 
the $\rho_{ab0}$ was defined as the low-temperature resistivity 
extrapolated to $T$\,=\,0. 
The Hall and magnetoresistance measurements were performed 
in a dilution refrigerator 
at stabilized temperature between 40\,mK and 1.6\,K, 
sweeping the magnetic field (applied\,$\parallel$\,$c$ axis) 
between $-12$\,T  and 12\,T, and deducing 
$R_{\rm H}$ from the gradient of the Hall resistivity $\rho_{\rm H}$ at low fields.  
$R_{\rm H}$ in all samples is almost temperature-independent below 1.6\,K. 

\section{Results and Discussion}

In Fig.~\ref{Fig1}, 
we show the Hall resistivity $\rho_{\rm H}$ for all the La-doped crystals that we have studied, 
along with an example from the Ti-doped system ($x$\,=\,0.01). 
In the Ti-doped sample, 
whose residual resistivity is 5\,$\mu\Omega$cm, 
$\omega_{\rm c}\tau$\,$\ll$\,1 over the whole field range, 
and the Hall resistivity remains field linear. 
This field linearity is shared by the La-doped crystals with high $y$ ($y$\,$>$\,0.10), 
for which $\rho_{ab0}$\,$>$\,3\,$\mu\Omega$cm. 
As $y$ and $\rho_{ab0}$ decrease, 
more and more curvature is seen in the Hall resistivity, 
due to increasing influence of the next order cubic term 
in the Zener-Jones expansion. 
All data for $y$\,$\geq$\,0.02 can be fitted very well 
by an expression of the form: 
\begin{eqnarray}
\rho_{\rm H} = R_{\rm H}B + R_{\rm 3}B^{3}
\label{eq2}.  
\end{eqnarray}
Here, 
$R_{\rm 3}$\,$\propto$\,$\tau^{2}$\,$\propto$\,$1/\rho_{ab0}^{2}$\,$\propto$\,
$1/(n_{\rm imp} \sin^{2}\delta_{0})^{2}$ 
as discussed in Eq.~(3) of ref. \onlinecite{Andy_Hall} 
($\delta_{0}$: phase shift, $n_{\rm imp}$: impurity concentration). 
For $x$\,=\,$y$\,=\,0, however, 
no fit based on a Zener-Jones expansion matches the data well 
over the entire field range. 
The reason for this is simple: 
in contrast to those used in the original Hall effect work 
with $\rho_{ab0}$\,=\,0.9\,$\mu\Omega$cm \cite{Andy_Hall}, 
the best modern crystals with $\rho_{ab0}$\,=\,90\,n$\Omega$cm 
are so pure ($\ell$\,=\,1.1\,$\mu$m) that the strong-field regime 
($\omega_{\rm c}\tau$\,$>$\,1) can be reached at fields as low as 4\,T. 
This high value of $\omega_{\rm c}\tau$ completely invalidates 
the use of any fitting based on the Zener-Jones expansion. 
Instead, 
we observe almost perfect linearity of $\rho_{\rm H}$ at high fields 
(shown as dashed line), 
with a value of $R_{\rm H}$ of $-2.1$\,$\times$\,$10^{-10}$\,m$^{3}$/C. 
In line with the results of classic measurements on pure elemental metals 
\cite{Hurd,Ashcroft}, 
this is in much better agreement with a calculation 
based on the full Fermi volume 
(i.e. the volume of the hole Fermi-Surface sheet minus 
that of the two electron sheets) 
than the weak-field value of $-0.71$\,$\times$\,$10^{-10}$\,m$^{3}$/C.
The fact that the Hall resistivity can be accounted for so well 
in terms of an analysis based on orbital effects 
already gives good evidence that such effects dominate 
the transport properties of doped Sr$_{2}$RuO$_{4}$. 
This assertion is further confirmed by analysis 
of the transverse magnetoresistance $\mathit{\Delta}\rho_{ab}/\rho_{ab}$, 
which we show in Fig.~\ref{Fig2}.  
Just as in the case of the Hall effect, 
the next term in the Zener-Jones expansion 
(a negative term in $(\omega_{\rm c}\tau)^{4}$) is seen 
to play an increasing role as $y$ is decreased. 
For $y$\,=\,0, 
saturation is seen in the magnetoresistance, 
and the data can be fitted very well with the simplest form of an expression 
for magnetoresistance going beyond the Zener-Jones expansion, 
namely that based on two fluids (one of electrons and one of holes) \cite{Tyler}:  
\begin{eqnarray}
\frac{\mathit{\Delta}\rho_{ab}}{\rho_{ab}} = \frac{p(\omega_{\rm c}\tau)^{2}}{1+q(\omega_{\rm c}\tau)^{2}}
\label{eq3}. 
\end{eqnarray}
The overall conclusion drawn from the analysis of the data shown 
in Figs.~\ref{Fig1} and \ref{Fig2} is that 
orbital effects mainly dominate the magnetotransport of Sr$_{2}$RuO$_{4}$. 
%

\begin{figure}
\includegraphics[width=65mm]{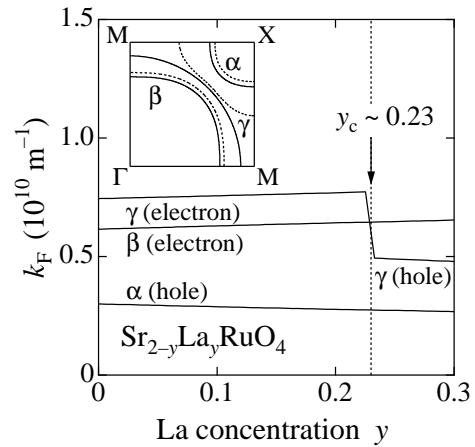}
\caption{\label{Fig3} 
Theoretical prediction of the dependence of 
the Fermi wave-vectors of the $\alpha$, $\beta$ and $\gamma$ bands 
on additional electron doping in Sr$_{2}$RuO$_{4}$, 
obtained using a tight-binding calculation. 
Note that the $\gamma$ Fermi surface changes from electron-like  
(solid lines in the inset) to hole-like  (dotted lines in the inset) 
when the Fermi energy moves beyond the van Hove singularity 
for $y$\,$>$\,0.23. 
In that case, the sign of the Hall coefficient 
$R_{\rm H}$ is predicted to change from negative to positive. }
\end{figure}

\begin{figure}
\includegraphics[width=75mm]{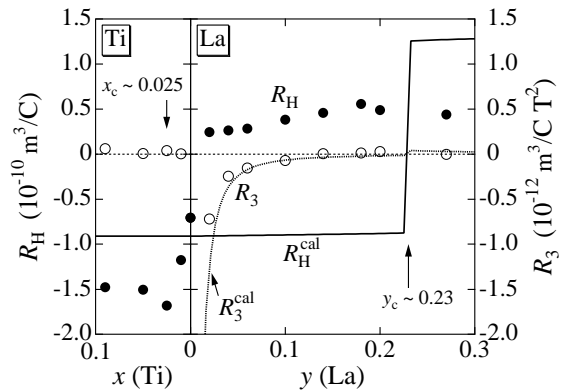}
\caption{\label{Fig4} 
The Hall coefficient $R_{\rm H}$ (closed circles) plotted against 
La and Ti concentration. 
The higher-order contribution $R_{\rm 3}$ (open circles) 
obtained by fitting using Eq.~(\ref{eq2}) is also plotted.  
The calculated $R_{\rm H}$ (denoted $R_{\rm H}^{\rm cal}$) 
is deduced from Eq.~(\ref{eq1}) using the same rigid-band model 
as that used for the calculations shown in Fig.~\ref{Fig2}. 
The calculated $R_{\rm 3}$ (denoted $R_{\rm 3}^{\rm cal}$) 
 is based on Eq.~(3) of ref. \onlinecite{Andy_Hall} (see text for details). 
 }
\end{figure}

%
In order to set the scene for the discussion of 
the weak-field Hall coefficient, 
we show in Fig.~\ref{Fig3} the rigid-band shift prediction 
for the evolution of $k_{\rm F}$ as a function of $y$ in 
Sr$_{2-y}$La$_{y}$RuO$_{4}$. 
Although these are predicted values, 
they are based on an empirically constrained band structure \cite{Bergemann_Review} and 
have been verified by direct measurement of quantum oscillations 
for $y$\,$\leq$\,0.06 \cite{Kiku_Rigid}. 
The main feature is the change in $k_{\rm F}$ for the $\gamma$ band 
at $y$\,=\,0.23, 
which corresponds to crossing of the van Hove singularity 
that lies approximately 49\,meV above the Fermi level 
in the undoped material \cite{CriticalPoint_Comment}. 
As shown in the inset to Fig.~\ref{Fig3}, 
the sharp change in the $k_{\rm F}$ of the $\gamma$ sheet coincides with 
its change from an electron pocket centered on $\Gamma$ 
to a hole pocket centered on X. 
Within the isotropic-$\ell$ approximation discussed in the introduction, 
this would lead to a sign reversal not just of the contribution of 
$\gamma$ band to $\sigma_{xy}$ but also of the overall Hall coefficient. 
The predictions for $R_{\rm H}$ deduced from Eq.~(\ref{eq1}) 
with ''isotropic-$\ell$'' model (denoted $R_{\rm H}^{\rm cal}$) 
are shown as a thin line in Fig.~\ref{Fig4} \cite{RH_Comment}. 
It is expected to be only very weakly $y$-dependent everywhere 
except the point at which $\gamma$ changes in character 
from electron-like to hole-like. 
The disagreement between the prediction of this simple model and the experimental data is clear. 
A sign reversal is seen in the data, but rather than being at the predicted value 
of $y$\,=\,0.23, it occurs centered on $y$\,$\sim$\,0.01.   
This observation serves as a topical warning against reading too much 
into simplified interpretations of the Hall effect in correlated electron metals.  
Not only does the sign change but the relative change in mod($R_{\rm H}$) is 
as large as a factor of three. 
All this occurs at a value of $y$ for which 
we have already obtained direct experimental proof that there is 
neither a large change in Fermi-surface topography 
nor a large change in the magnetic behavior of the system. 
There is no evidence at all for a quantum critical point at $y$\,$\sim$\,0.01.  
In constructing simplified models for the analysis of the Hall effect, 
two approximations are most commonly used. 
We have adopted the ''isotropic-$\ell$'' approach in which 
the assumption is that $\ell$\,=\,$v_{\rm F}\tau$ 
($v_{\rm F}$: Fermi velocity) is the essential quantity 
in determining transport properties 
if impurity scattering dominates the relaxation processes, i.e. that 
$v_{\rm F}$ and $\tau$ are no longer independent in this regime. 
An alternative that is also sometimes used is to factorize 
into $v_{\rm F}$ and $\tau$ and then to assume that 
$\tau$ is independent of $k$ in performing the transport integrals \cite{Norman}.  
In our opinion this latter approach is harder to justify 
when analyzing low temperature data, 
but it is also important to point out that its adoption would not account for 
our observation, 
because the $y$ dependence of the Fermi velocities is known to be very weak 
in this range of doping \cite{Kiku_Rigid}.
%

%
The results for La doping came as a considerable surprise, 
because the simple analysis had worked well for both undoped crystals 
and crystals containing Al impurities \cite{Galvin}. 
In order to obtain a better overall experimental picture of the effect of doping, 
we have also studied a series of crystals in which 
non-magnetic Ti$^{4+}$ has been substituted for the in-plane Ru$^{4+}$. 
The results are also shown in Fig.~\ref{Fig4}.  
Although there is variation in mod($R_{\rm H}$), 
there is neither a sharp step nor a sign change \cite{RH_at_QCP_Comment}.  
The interpretation that can be drawn 
within an orbital analysis of the Hall effect is that 
each substitution has some band- or even $k$-specific effect 
on the mean free path, 
invalidating the use of the isotropic-$\ell$ approximation. 
In fact, 
we can quantitatively model the $y$ dependence of parameter $R_{\rm 3}$ 
from Eq.~(\ref{eq2}) by making the assumption that in the La-doped samples, 
the mean-free path for the holes is a factor of 1.6 longer than that for the electrons. 
The calculated $R_{\rm 3}$ (denoted $R_{\rm 3}^{\rm cal}$) is shown 
in Fig.~\ref{Fig4}. 
As discussed in ref. \onlinecite{Andy_Hall}, 
this small factor of 1.6 would be enough to explain 
the the experimental $R_{\rm H}$ 
with $\sim$\,$+0.4$\,$\times$\,$10^{-10}$\,m$^{3}$/C  
if the onset of sign change coincided with the substitution of La into the crystals, 
and would also contribute to the violation of ''Kohler's rule'' 
seen in the magnetoresistance data of Fig.~\ref{Fig2}. 
The analysis of this kind is simplified, 
because it ignores the possibility of differences in $\ell$ 
between the $\beta$ and $\gamma$ sheets, 
and $k$-dependent changes within any one sheet. 
However, it strongly suggests that 
deviations of some kind from the isotropic-$\ell$ approximation are 
responsible for the observed differences in the effects of doping 
with La or Ti (this work) or Al. 
For Sr$_{2-y}$La$_{y}$RuO$_{4}$, 
this might be due to the unique role of La 
in band-selectively increasing the magnetic fluctuations on $\gamma$ sheet. 
As shown in refs. \onlinecite{Kiku_Rigid} and \onlinecite{Kiku_Band}, 
the fact that La doping shifts the Fermi level towards the van Hove singularity 
on the $\gamma$ sheet leads to a change in both its specific heat and low-$q$ magnetic susceptibility. 
In contrast, 
the dominant effect of Ti is to enhance the well-known 
incommensurate fluctuations \cite{Sidis} at the wave vector 
$\textit{\textbf{q}}$\,=\,$\textit{\textbf{Q}}_{\rm ic}^{{\alpha}{\beta}}$\,$\sim$\,(2$\pi$/3,\,2$\pi$/3,\,0) toward $x$\,$\sim$\,0.025, 
eventually producing spin-density-wave (SDW) order for $x$\,$\geq$\,0.025 
\cite{Braden_Ti}. 
This effect is known to be associated with the nesting vector between  
the $\alpha$ and $\beta$ Fermi surfaces \cite{Mazin_PRL99}. 
It should be noted here that 
the dopant of Al is not thought to have any pronounced effect 
on the magnetic properties \cite{Galvin}. 
Although it might eventually be possible to construct a model for $R_{\rm H}$ 
based on detailed knowledge of $k$-dependent scattering in Sr$_{2}$RuO$_{4}$, 
it would probably end up with so many free parameters 
that it would be difficult to make a firm judgement on its validity. 
That is probably the main message from the work reported here. 
Although the weak-field Hall effect is often referred to as a simple probe of 
Fermi-surface topography, 
its simplicity relates more to the ease with which 
it can be measured than that with which it can be interpreted. 
Even in the low-temperature limit in which its relationship to 
Fermi-surface topography should be at its clearest, 
its results can be seriously misleading.  
Here, we have given a concrete example of a sudden jump 
in the value of $R_{\rm H}$ as a function of doping 
which definitely does not signal a correspondingly large change 
in Fermi-surface topography. 
%

\section{Summary}

We have presented detailed measurements of the Hall effect 
in La- and Ti-doped single-crystals of Sr$_{2}$RuO$_{4}$. 
Analysis of data taken to high fields argues strongly 
in favor of an interpretation based dominantly on the orbital Hall effect. 
In spite of this, we report a pronounced anomaly as a function of La doping 
that is not simply related to a change in the Fermi surface. 
%

\section*{Acknowledgments} 

The authors thank A.J. Schofield, M. Sing and S.R. Julian 
for useful discussions. 
They also thank R.A. Borzi and S.A. Grigera 
for technical supports and useful comments, 
Y. Shibata, J. Hori, Takashi Suzuki, and Toshizo Fujita 
for electron-probe micro-analysis (EPMA) 
measurements at Hiroshima University. 
This work was supported 
by the Grant-in-Aid for Scientific Research 
(S) from the Japan Society for Promotion of Science (JSPS), by the Grant-in-Aid 
for Scientific Research on Priority Area 'Novel Quantum Phenomena in 
Transition Metal Oxides' from the Ministry of Education, Culture, Sports, 
Science and Technology.
One of the authors (N.K.) is supported by JSPS Postdoctoral Fellowships 
for Research Abroad.  
%



\begin{thebibliography}{99}
\bibitem{CriticalPoint_Comment}
According to ref.~\onlinecite{Kiku_Band}, 
the instability point occurs for $y$\,$\sim$\,0.20. 
%
The difference from $y$\,=\,0.23 is probably 
because the decrease of 
the $c$-axis lattice parameter with La doping, 
which lowers the $\gamma$ band relative to the others, 
might not be negligible for higher concentration.
\bibitem{RH_Comment}
For simplicity, 
$R_{\rm H}^{\rm cal}$ in Sr$_{2}$Ru$_{1-x}$Ti$_{x}$O$_{4}$ assumes 
that the Fermi wave-vector is unchanged with $x$ for all bands, 
although a (tiny) SDW gap should be opened for $x$\,$\geq$\,0.025 
\cite{Braden_Ti,Sing}.
\bibitem{RH_at_QCP_Comment}
We note that the $R_{\rm H}$ as a function of both La and Ti concentrations 
shows cusps corresponding to each magnetic instability point 
at $y$\,$\sim$\,0.20 \cite{Kiku_Band} and $x$\,$\sim$\,0.025 \cite{Kiku_Ti}.
\end{thebibliography}
\end{document}